\documentclass[tightenlines,superscriptaddress,eqsecnum,floats,nofootinbib,showpacs]{revtex4-2}
\usepackage{amssymb}
\usepackage{stmaryrd}
\usepackage{amsmath}
\usepackage{amsfonts}
\usepackage{mathrsfs}
\usepackage{amsmath,amssymb,amsfonts}
\usepackage[utf8]{inputenc}
\usepackage{graphicx}
\usepackage{subfigure}
\usepackage{color} 
\def\be{\begin{equation}}
	\def\ee{\end{equation}}
\def\ba{\begin{eqnarray}}
	\def\ea{\end{eqnarray}}

\newcommand{\eqnref}[1]{(\ref{#1})}

\begin{document}

	\title{Bosonic and Fermionic love number of static acoustic black hole}
	\author{Yongbin Du}
	\affiliation{School of Physics and Astronomy, Sun Yat-sen University, Zhuhai, 519082, China}


	\author{Xiangdong Zhang\footnote{Corresponding author. scxdzhang@scut.edu.cn}}
	\affiliation{School of Physics and Optoelectronics, South China University of Technology, Guangzhou 510641, China}
	\date{\today}

	\begin{abstract}
We compute static ($\omega\to0$) tilde Love numbers for scalar ($s=0$) and Dirac ($s=1/2$) perturbations of static acoustic black holes (ABHs) in (3+1) and (2+1) dimensions respectively. By imposing horizon regularity condition and matching to the large-radius expansion, we extract the ratio between decaying and growing modes. It turns out that in (3+1) dimensions the scalar Love number is generically nonzero for ABHs, while the Fermionic Love numbers follow a universal power-law form $F^{\pm1/2}_{\ell m}=\pm 4^{-(\ell+1/2)}$. In (2+1) dimensions the scalar field exhibits a strange logarithmic structure, causing the Bosonic Love number to vanish for even $m$ but remain nontrivial for odd $m$; In contrast, the Fermionic Love number in this case retains a simple power-law form $F_m=4^{-m}$ and is generically nonzero. These results provide insights into tidal response in analogue gravity systems and highlight qualitative differences between integer- and half-integer-spin fields.
\end{abstract}

	\maketitle

\newpage
\tableofcontents

	\section{Introduction}
When a compact object is subjected to an external tidal field, it undergoes a tidal deformation that, in turn, alters the multipolar structure of the surrounding field. This induced multipolar response is parametrized by the Love numbers (LNs) \cite{BinningtonPoisson:2009,DamourNagar:2009}. These quantities can be interpreted as linear-response coefficients of the object to time-independent external sources; in the frequency domain, this corresponds to the zero-frequency limit of the retarded Green's function. 
Since the external sources may excite scalar ($s=0$), spinor ($s=1/2$), vector ($s=1$), or even gravitational ($s=2$) fields, Love numbers can be classified accordingly, characterizing the zero-frequency linear response of the object to different field types.
In general, a physical object exhibits a non-vanishing response to an external tidal field, which can be understood as the inevitable development of a deformation induced by the field. Objects that do not deform under external perturbations are referred to as ``rigid bodies". Much like the concept of a black body, rigid bodies have long been regarded as idealized, Platonic constructs rather than physically realizable entities. Remarkably, the advent of black holes in general relativity challenges this conventional perspective.

It has been demonstrated that, for four-dimensional spherically symmetric black holes in general relativity, the Love numbers ‌vanish identically in the scalar, vector, and gravitational sectors. \cite{BinningtonPoisson:2009,DamourNagar:2009,FangLovelace:2005,HuiJoycePencoSantoniSolomon:2021}.
The same property persists for rotating and electrically charged solutions \cite{CharalambousDubovskyIvanov:2021,PereniguezCardoso:2022}.
However, the vanishing of Love numbers for Bosonic fields with integer spin no longer holds in higher-dimensional spacetimes \cite{KolSmolkin:2012,CardosoGualtieriMoore:2019}, in asymptotically AdS backgrounds \cite{EmparanFernandezPiqueLuna:2017,FranzinFrassinoRocha:2024}, or within modified theories of gravity \cite{CardosoKimuraMaselliSenatore:2018,CardosoFranzinMaselliPaniRaposo:2017,Liu2025Love}. This suggests that the vanishing of Love numbers appears to be a distinctive feature of general relativity.

However, recent studies have demonstrated that, when considering Dirac spinor fields around Schwarzschild and Kerr black holes \cite{ChakrabortyHeidmannPani:2025}, the corresponding Fermionic Love numbers no longer vanish. Subsequent analyses of Reissner–Nordström black holes have corroborated this behavior \cite{PangTianZhangJiang:2025}.
This implies that the vanishing of Love numbers is not an inevitable consequence of general relativity. Instead, the question of whether Love numbers vanish may point to a deeper underlying mechanism. One proposed explanation attributes this phenomenon to symmetry considerations \cite{CharalambousDubovskyIvanov:2021,HuiJoycePencoSantoniSolomon:2022,RaiSantoni:2024,SharmaGhoshSarkar:2024}.  In curved spacetimes, Love numbers are typically defined as the ratio between the coefficients of decaying and growing modes in the asymptotic expansion of field solutions at spatial infinity. When the differential operator governing the field equations can be identified with the Casimir operator of a symmetry group, for instance $SL(2,\mathbb{C})$, the theory inherits the corresponding symmetry, which imposes strong constraints on the solution structure. Then the vanishing of Bosonic Love number is explained as caused by the vanishing of decay solution branch, arise from the symmetry of the theory.

While symmetry-based arguments provide an insightful explanation for the vanishing of Love numbers, they function more as a heuristic tool for estimating these quantities for a given field equation rather than as a fundamental explanation for their disappearance.
To gain further insight, it is desirable to accumulate explicit computations of Love numbers across a broader range of backgrounds. A particularly meaningful arena for such investigations is provided by analogue gravity systems \cite{Visser:1998,BarceloLiberatiVisser:2005}.
For instance, ref.~\cite{Unruh1981} constructed an acoustic black hole (ABH), where the equations governing an incompressible fluid flow in a duct take the form of a field theory in curved spacetime. In this setup, sound-wave perturbations serve as analogues of matter fields propagating in curved backgrounds. The system admits an acoustic horizon, which plays a role analogous to the event horizon in general relativity. Acoustic black holes have been successfully employed to simulate phenomena such as superradiance \cite{BasakMajumdar2003}, Hawking radiation \cite{VieiraBezerra2016,VieiraBezerra2019Erratum}, and quasinormal modes \cite{BertiCardosoLemos2004,Saavedra2006}.
Some later works have shown that acoustic black holes can be embedded into the fluid effective theory framework of holography \cite{Ge:2015AcousticHolography,SunYu:2019D3Acoustic}. More recently, some interesting work find that the island prescription can be applied to an acoustic black hole in a Schwarzschild background \cite{ChengSun:2025IslandABH}.

Although analogue black holes can reproduce a wide range of gravitational phenomena, their physical nature fundamentally differs from that of black holes in general relativity: acoustic black holes are composed of ordinary matter and thus do not constitute ``rigid bodies" in the strict sense. It is therefore a nontrivial and intriguing question whether, under scalar and spinor perturbations, they exhibit Love-number properties analogous to those of general-relativistic black holes. While Ref.~\cite{DeLucaKhekKhouryTrodden2025} investigated scalar Love numbers in such systems. In the present work, we extend this line of research by studying both the Bosonic Love numbers associated with $s=0$ scalar fields and the Fermionic Love numbers associated with $s=1/2$ spinor fields in four- and three-dimensional static acoustic black-hole backgrounds. It allows us to examine in a sharp manner which properties of black holes can be faithfully simulated and which cannot. Not only of theoretical interest, it is also potentially valuable from an experimental perspective: if one could engineer an analogue-gravity setup that effectively realizes spinor fields in curved spacetime, then Fermionic Love numbers would become, in principle, measurable in the laboratory, with potentially far-reaching implications for our understanding of real black holes.

This paper is organized as follow: In section~2, we compute the Love numbers for scalar and spinor fields in four-dimensional acoustic black-hole backgrounds. In section~3, we further investigate how these results are modified in one lower dimension, namely for three-dimensional acoustic black holes, focusing on both the Bosonic Love numbers of scalar fields and the Fermionic Love numbers of spinor fields. Finally, in section~4 we summarize our results and present a brief outlook.


\section{Love number of (3+1) dimensional ABH}

\subsection{Scalar field}

In $(3+1)$ dimensions, the most general spherically symmetric acoustic black-hole metric can be written as \cite{Visser:1998}
\begin{equation}\label{4dim}
ds^{2}=- c_s^{2}\, f(r)\, dt^{2}+\frac{dr^{2}}{f(r)}+r^{2}\left(d\theta^{2}+\sin^{2}\theta\,d\phi^{2}\right),
\end{equation}
where $c_s$ denotes the (effective) sound speed and $f(r)=1-(r_{+}/r)^4$ is a radial function encoding the presence of an acoustic horizon at $f(r_+)=0$.
We construct a massless scalar field propagating on this background. So the equation of motion is the Klein-Gordon form, written as
\begin{equation}
-\frac{1}{f(r)c_s^{\,2}}\frac{\partial^2 \Phi}{\partial t^2}
+\frac{1}{r^2}\frac{\partial}{\partial r}\!\left(r^2 f(r)\frac{\partial \Phi}{\partial r}\right)
+\frac{1}{r^2\sin\theta}\frac{\partial}{\partial \theta}\!\left(\sin\theta\,\frac{\partial \Phi}{\partial \theta}\right)
+\frac{1}{r^2\sin^2\theta}\frac{\partial^2 \Phi}{\partial \phi^2}
=0.
\end{equation}
We separate variables as $\Phi(t,r,\theta,\phi)=e^{-i\omega t}R(r)Y(\theta,\phi)$, The angular dependence is governed by the spherical harmonics $Y(\theta,\phi)=Y_{\ell m}(\theta,\phi)$,
which satisfy the eigenvalue equation on the unit two-sphere,
\begin{equation}
\frac{1}{\sin\theta}\frac{\partial}{\partial\theta}\!\left(\sin\theta\,\frac{\partial Y}{\partial\theta}\right)+\frac{1}{\sin^2\theta}\frac{\partial^2 Y}{\partial\phi^2}+ \ell(\ell+1)\,Y 
= 0 ,
\end{equation}
where $\ell=0,1,2\,...$.Substituting it into the Klein-Gordon equation, one finds that the radial function $R(r)$ satisfies
\begin{eqnarray}
\frac{1}{r^2}\frac{d}{dr}\!\left(r^2 f(r)\frac{dR}{dr}\right)
+\left(\frac{\omega^2}{f(r)c_s^{\,2}}-\frac{\ell(\ell+1)}{r^2}\right)R &=& 0.
\end{eqnarray}
Next, to solve this field equation in the zero-frequency limit $\omega=0$, since the function $f(r)$ contains a quartic term in $r$, we introduce the following change of variables 
\begin{equation}
z = 1 - \frac{r^{4}}{r_{+}^{4}}.
\end{equation}
In this way, the radial function $R(z)$ satisfies the equation
\begin{equation}
z(1-z)\frac{d^2 R}{dz^2}
+(1-\frac{5}{4}z)\frac{dR}{dz}
+\frac{\ell(\ell+1)}{16}R=0.
\end{equation}
This is a hyper-geometric equation with $a=(\ell+1)/4$,\,\,$b=-\ell/4$,\,\,$c=1$.
The equation has three regular singularities at $z=1,0,-\infty$ respectively. Here $z=0$ corresponds to the acoustic horizon. Near this regular singular point, the general local solution is typically expressed as a linear combination of two linearly independent hyper-geometric solutions \cite{hyper}.
However, in the present case one has $c=1\in \mathbb{Z}$, so one of the fundamental solutions in the neighbourhood of $z=0$ develops a logarithmic behavior of the form
\begin{equation}
R_{2}(z)\sim (\text{const.})\; {}_2F_1(a,b;c;z)\,\ln z \;+\; (\text{power series}),
\end{equation}
where ${}_2F_1(a,b;c;z)$ is Gauss hyper-geometric function.
This solution is manifestly non-regular at the horizon $z=0$ and must therefore be discarded.
Consequently, the unique physically acceptable solution in the near-horizon region is given by
\begin{equation}
R_1(z) = {}_2F_1\!\left(\frac{\ell+1}{4},\,-\frac{\ell}{4};\,1;\,z\right).
\end{equation}
Using the Kummer solutions together with the corresponding connection coefficients, the above solution can be expanded in the asymptotic region $z\to\infty$ as follows \cite{hyper}
\begin{eqnarray}\label{expanding}
 {}_2F_1(a,b;c;z) &=&
\frac{\Gamma(c)\Gamma(b-a)}{\Gamma(b)\Gamma(c-a)}
(-z)^{-a}{}_2F_1\!\left(a,1-c+a;1-b+a;\frac{1}{z}\right)
\\\notag
&+&\frac{\Gamma(c)\Gamma(a-b)}{\Gamma(a)\Gamma(c-b)}(-z)^{-b}
{}_2F_1\!\left(b,1-c+b;1-a+b;\frac{1}{z}\right) .
\end{eqnarray}
So in the limit $z\to \infty$, the radial function behave as 
\begin{equation}
R_1(r)\sim A\,\,(\frac{r}{r_{+}})^{-\ell-1}+B\,\,(\frac{r}{r_{+}})^{\ell},
\qquad (r\to\infty).
\end{equation}
at the infinity. The coefficiencts of the growing modes and decay mode are
\begin{equation}
A=\frac{\Gamma\!\left(-\frac{2\ell+1}{4}\right)}
{\Gamma\!\left(-\frac{\ell}{4}\right)\Gamma\!\left(\frac{3-\ell}{4}\right)},
\qquad
B=\frac{\Gamma\!\left(\frac{2\ell+1}{4}\right)}
{\Gamma\!\left(\frac{\ell+1}{4}\right)\Gamma\!\left(1+\frac{\ell}{4}\right)}.
\end{equation}
Love number is given by the ratio between the coefficient of the decaying mode and that of the growing mode of the zero-frequency solution in the asymptotic region, which is a dimensionless parameter defined as
\begin{equation}
    \mathcal{F}_{\ell m}=\frac{\text{coefficiencts of decay modes}}{\text{coefficiencts of growing modes}}=\frac{A}{B}
\end{equation}

Thus we obtain
\begin{equation}
\mathcal{F}_{\ell m}=
-\frac{1}{\pi^2}\sin\!\left(\frac{\pi\ell}{4}\right)\,
\sin\!\left(\frac{\pi(\ell+1)}{4}\right)\,
\Gamma^2\!\left(\frac{\ell+1}{4}\right)\,
\frac{\Gamma\!\left(-\frac{2\ell+1}{4}\right)}{\Gamma\!\left(\frac{2\ell+1}{4}\right)}.
\end{equation}
For certain special values of $\ell$, e.g. when $\ell$ is a integer multiple of $4$, the Love number can vanish. But in general, in sharp contrast to the Schwarzschild case, the Love number for scalar field of an acoustic black hole is nonzero and its expression is complicated.
This result resonates with our original motivation: as an analogue black hole made of ordinary matter, an acoustic black hole may not have to inherit the property of having a vanishing static response in the scalar field.

\subsection{Spinor field}
Now we proceed to assume that a analogue Dirac spinor field propagates on the background \eqnref{4dim}. In flat spacetime, the spinor obeys Dirac equation
\begin{equation}
i \gamma^{\mu} \partial_{\mu} \Psi - m \Psi = 0 .
\end{equation}
In the massless case considered in this work, the Dirac equation decouples into left-handed and right-handed Weyl equations,
\begin{equation}
\sigma^{\mu} \partial_{\mu} \psi_L = 0 ,
\qquad
\bar{\sigma}^{\mu} \partial_{\mu} \psi_R = 0 ,
\end{equation}
where $\sigma^{\mu}=(I,\sigma_{x},\sigma_{y},\sigma_{z})$ and $\bar{\sigma}^{\mu}=(I,-\sigma_{x},-\sigma_{y},-\sigma_{z})$ are vectors whose components are $2 \times 2$ identity matrix $I$ and Pauli matrices.
These equations are originally defined in flat spacetime. When the 4- dimensional background geometry is curved, they can be reformulated using the Newman-Penrose (NP) formalism \cite{Chandrasekhar:1983mtbh}. 
To facilitate a more compact construction of the Newman-Penrose tetrad, we express the spherically symmetric metric in Eddington-Finkelstein coordinates $(v,r,\theta,\phi)$ as
\begin{eqnarray}
    ds^2=- f(r)c_{s}^2\, dv^2+2c_{s}dvdr+ r^2 d\theta^2+ r^2 \sin^2\theta\, d\phi^2. 
\end{eqnarray}
where $f(r)=1-(r_{+}/r)^4$. For this form of the metric, the Newman-Penrose null tetrad can be chosen as

\begin{eqnarray}
l^{\mu}&=&\left(1,\;\frac{c_s f(r)}{2},\;0,\;0\right),\qquad
n^{\mu}=\left(0,\;-\frac{1}{c_s},\;0,\;0\right),\\\notag
m^{\mu}&=&\frac{1}{\sqrt{2}\,r}\left(0,\;0,\;1,\;\frac{i}{\sin\theta}\right),\qquad
\bar m^{\mu}=\frac{1}{\sqrt{2}\,r}
\left(0,\;0,\;1,\;-\frac{i}{\sin\theta}\right).
\end{eqnarray}
Using the above tetrad, one finds the following nonzero complex spin coefficients
\begin{eqnarray}
\rho=-\frac{c_{s}f(r)}{2r},\qquad
\mu=-\frac{1}{rc_{s}},\qquad
\epsilon=\frac{f'(r)c_s}{4},\qquad
\alpha=-\frac{\cot\theta}{2\sqrt{2}\,r},\qquad
\beta=\frac{\cot\theta}{2\sqrt{2}\,r},
\end{eqnarray}
Without loss of generality, we restrict our attention to the left-handed Weyl equation, the form of which written by spin coefficients is \cite{Chandrasekhar:1983mtbh}
\begin{eqnarray} (D+\epsilon-\rho)\,F_1+(\bar\delta+\pi-\alpha)\,F_2 &=& 0,\\\notag
(\Delta+\mu-\gamma)\,F_2+(\delta+\beta-\tau)\,F_1 &=& 0, \end{eqnarray}
where $D=l^{\mu}\partial_{\mu}$,\,$\Delta=n^{\mu}\partial_{\mu}$,\,$\delta=m^{\mu}\partial_{\mu}$,\,$\bar{\delta}=\bar{m}^{\mu}\partial_{\mu}$ and $F_{1}(x)$, $F_{2}(x)$ represent the components of the spinor field $\psi^{A}_{L}=F_{1}\, o^{A}+F_{2}\, l^{A}$ in dyads $(o^{A},\, l^{A})$, which satisfy
\begin{eqnarray}
o_{A} o^{A}=l_{A} l^{A}=0,
\qquad
o_{A} l^{A}=-\,o^{A} l_{A}=1 
\end{eqnarray}
So we can write the equation as 
\begin{eqnarray}
\Bigl(\partial_v+c_{s}\frac{f}{2}\partial_r+c_{s}\frac{f'}{4}+c_{s}\frac{f}{2r}\Bigr)F_1
+\Bigl(\frac1{\sqrt2\,r}(\partial_\theta-i\csc\theta\,\partial_\phi)+\frac{\cot\theta}{2\sqrt2\,r}\Bigr)F_2
&=&0,
\\[8pt]
\Bigl(-\frac{1}{c_{s}}\partial_r-\frac{1}{c_{s}r}\Bigr)F_2
+\Bigl(\frac1{\sqrt2\,r}(\partial_\theta+i\csc\theta\,\partial_\phi)+\frac{\cot\theta}{2\sqrt2\,r}\Bigr)F_1
&=&0 .
\end{eqnarray}

Next, we adopt the method of separation of variables and write $F(x)$ in the following form
\begin{eqnarray}
    F_1(v,r,\theta,\phi)=e^{-i\omega v}\,R_{-}(r)\,S_{-}(\theta,\phi),\qquad
F_2(v,r,\theta,\phi)=e^{-i\omega v}\,R_{+}(r)\,S_{+}(\theta,\phi).
\end{eqnarray}
 angular equations
\begin{eqnarray} 
\mathcal L_- S_+ &=& \lambda\, S_-,\\[4pt] \mathcal L_+ S_- &=& -\lambda\, S_+,
\end{eqnarray}
where
\begin{eqnarray}
    \mathcal{L}_{+}\equiv \partial_\theta+i\,\csc\theta\,\partial_\phi+\frac{1}{2}\cot\theta,\qquad
\mathcal{L}_{-}\equiv \partial_\theta-i\,\csc\theta\,\partial_\phi+\frac{1}{2}\cot\theta.
\end{eqnarray}
It is evident that $S_{\pm}(\theta,\phi)$ corresponds to the spherical harmonics with spin weight $s=\pm\frac{1}{2}$
\begin{eqnarray}
    S_{-}(\theta,\phi)= {}_{-\frac12}Y_{\ell m}(\theta,\phi),\qquad
S_{+}(\theta,\phi)= {}_{+\frac12}Y_{\ell m}(\theta,\phi),
\end{eqnarray}
with the eigenvalue
\begin{eqnarray}
    \lambda=\ell+\frac12,\qquad
\ell=\frac12,\frac32,\frac52,\cdots,\qquad
m=-\ell,-\ell+1,\cdots,\ell.
\end{eqnarray}
Now we set
\begin{eqnarray}\label{uR}
u_{-}(r)\equiv r\,R_{-}(r),\qquad
u_{+}(r)\equiv r\,R_{+}(r)
\end{eqnarray}
and $\omega\to0$ to rewrite the radial equation
\begin{eqnarray} \label{radialeom1}
c_{s}\frac{f}{2}\frac{du_-}{dr}+c_{s}\frac{f'}{4}u_-+\frac{\lambda}{\sqrt2\,r}u_+&=&0,\\\notag \frac{du_+}{dr}+c_{s}\frac{\lambda}{\sqrt2\,r}u_-&=&0, 
\end{eqnarray}
so we can eliminate $u_{+}$ easily and thereby obtain the negative spin branch
\begin{eqnarray}\label{radialeom2}
    r f\,u_{-}''+\left(f+\frac{3}{2}r f'\right)u_{-}'
+\left(\frac{f'}{2}+\frac{r f''}{2}-\frac{\lambda^{2}}{r}\right)u_{-}=0.
\end{eqnarray}
So the radial equation is
\begin{eqnarray}
r^{2} f\, R_-'' +\left( 3 r f + \frac{3}{2} r^{2} f' \right) R_-'
+\left( f + 2 r f' + \frac{r^{2}}{2} f'' - \lambda^{2} \right) R_- \\\notag
=\left(r^{2}-\frac{r_{+}^{4}}{r^{2}}\right) R_-'' +\left(3r+\frac{3 r_{+}^{4}}{r^{3}}\right) R_-'+\left(
1-\lambda^{2}-\frac{3 r_{+}^{4}}{r^{4}}\right) R_-=0 .
\end{eqnarray}
Let $z=1-r^{4}/r_{+}^{4}$ again, the equation becomes
\begin{equation}
-16\,z(1-z)\,R_{zz}
+\left(24z-24\right)R_{z}
+\left(1-\lambda^{2}-\frac{3}{1-z}\right)R=0,
\end{equation}
where $R_{z}\equiv dR/dz$ and $R_{zz}\equiv d^{2}R/dz^{2}$. In order to rewrite this equation into a more tractable form, we further make the substitution $R(z)=(1-z)^{\beta}y(z)$, where $\beta=1/4$. After this substitution, the equation takes the form
\begin{equation}
z(1-z)\,y''(z)
+\left(\frac{3}{2}-2z\right)y'(z)
-\left(\frac{1}{4}-\frac{\lambda^{2}}{16}\right)y(z)
=0.
\end{equation}
This is precisely the standard hyper-geometric equation, with parameters
\begin{equation}
a=\frac{1}{2}+\frac{\lambda}{4},\qquad
b=\frac{1}{2}-\frac{\lambda}{4},\qquad
c=\frac{3}{2}.
\end{equation}
As $z\to 0$ (near the horizon), the regular branch of solutions is
\begin{equation}
y_{1}(z) = {}_2F_1(a,b;c;z)
= {}_2F_1\!\left(\frac{1}{2}+\frac{\lambda}{4},\,\frac{1}{2}-\frac{\lambda}{4};\,\frac{3}{2};\,z\right).
\end{equation}
Since $1-a=b$ and $c=3/2$, this hyper-geometric function admits a representation in terms of elementary functions using Eq.~(15.4.15) in Ref.~\cite{hyper}
\begin{equation}
    {}_{2}F_{1}(a,1-a,\frac{3}{2},-x^2)=\frac{1}{(2-4a)x}\Big[(\sqrt{1+x^2}+x)^{1-2a}-(\sqrt{1+x^2}-x)^{1-2a}\Big]
\end{equation}
When replacing $x^2\to-z$, the regular solution $y_{1}(z)$ at infinity behaves as
\begin{eqnarray}
    y_{1}(z)\to A_{\infty}(-z)^{-\frac{\lambda+1}{2}}+B_{\infty}(-z)^{\frac{\lambda-1}{2}}
    \to A_{\infty}(\frac{r}{r_{+}})^{-2(\lambda+1)}+B_{\infty}(\frac{r}{r_{+}})^{2(\lambda-1)}
\end{eqnarray}
where
\begin{eqnarray}
    A_{\infty}=-\frac{2^{-\lambda-1}}{\lambda},\qquad B_{\infty}=\frac{2^{\lambda-1}}{\lambda}
\end{eqnarray}
Consequently, we obtain the Fermionic Love number with $s=-1/2$ as 
\begin{eqnarray}
    \mathcal{F}_{-\frac{1}{2}\ell m}=\frac{A_{\infty}}{B_{\infty}}=-\frac{1}{4^{\lambda}}.
\end{eqnarray}
Taking into account the result for the positive branch $R_{+}$ corresponding to $s=1/2$ as well, we finally find that the Fermionic Love number for a spherically symmetric acoustic black hole takes a form similar to the Schwarzschild case, namely,
\begin{eqnarray}
    \mathcal{F}_{\pm\frac{1}{2}\ell m}=\pm\frac{1}{4^{\ell+\frac{1}{2}}}.
\end{eqnarray}
We therefore find that, although the scalar Love numbers take a rather intricate form, the Fermionic Love numbers of the acoustic black hole exhibit a structure closely analogous to the Schwarzschild result $\pm4^{-(2\ell+1)}$ \cite{ChakrabortyHeidmannPani:2025}. Unlike the Bosonic case, the Fermionic Love numbers do not possess any zeros. This feature appears to be universal across a variety of gravitational models, suggesting that spinor fields display a qualitatively different behavior from scalar fields.

\section{Love number of (2+1) dimensional ABH}
\subsection{Scalar field}
In the following, we focus on a $(2+1)$-dimensional acoustic black hole in the nonrotating limit. The metric reads \cite{Visser:1998}
\begin{equation}\label{3dimmetric}
ds^{2}=- c_s^{\,2} f(r)\,dt^{2}+\frac{dr^{2}}{f(r)}+r^{2} d\phi^{2},
\end{equation}
where $f(r)=1-(r_{+}/r)^2$. $r=r_{+}=A/c_{s}$ is a circle representing a codimension-2 acoustic horizon. $A$ is the radial component of the velocity field \cite{Visser:1998}. The function $f(r)$ is formally similar to that of higher-dimensional Schwarzschild black holes.

Now we consider an effective scalar field propagating on this background.
The equation of motion for the scalar field, for instance, effective phonon field, in the massless and minimally coupled limit, reads
\begin{equation}
-\frac{1}{c_s^{2} f(r)}\,\partial_t^{2}\Phi
+\frac{1}{r}\,\partial_r\!\left(r\,f(r)\,\partial_r\Phi\right)
+\frac{1}{r^{2}}\,\partial_\phi^{2}\Phi=0.
\end{equation}
We adopt the separation of variables ansatz
\begin{equation}
\Phi(t,r,\phi)=e^{-i\omega t}\,R(r)\,\Theta(\phi).
\end{equation}
Substituting this ansatz into the Klein-Gordon equation in its most general form, we obtain the corresponding radial and angular equations.
The angular equation takes the form
\begin{equation}
\Theta''(\phi)+m^{2}\Theta(\phi)=0.
\end{equation}
Imposing the periodic boundary condition $\Theta(\phi+2\pi)=\Theta(\phi)$, the normalized angular eigenfunctions are given by
\begin{equation}
\Theta_m(\phi)=e^{i m \phi},
\qquad
m\in\mathbb{Z}.
\end{equation}
Without loss of generality, we take $m\ge 0$ in the following. In the zero-frequency limit, the radial equation reads
\begin{eqnarray}
&&\frac{1}{r}\frac{d}{dr}\left(r\,f(r)\,\frac{dR}{dr}\right)
-\frac{m^{2}}{r^{2}}
R(r)\\\notag
&=&\left(r^{2}-r_{+}^{2}\right) R''(r)
+\left(r+\frac{r_{+}^{2}}{r}\right) R'(r)- m^{2} R(r)=0 .
\end{eqnarray}
With the change of variable $z=1-\left(r/r_{+}\right)^2$, we obtain the differential equation for $R(z)$
\begin{equation}
z(1-z)\,R''(z)+(1-z)\,R'(z)+\frac{m^{2}}{4}\,R(z)=0 .
\end{equation}
This is a hyper-geometrical equation with $a=-m/2$,\,$b=m/2$ and $c=1$. Its regular branch of solution at horizon $z=0$ is
\begin{eqnarray}
R_{1}(z)={}_2F_{1}(-\frac{m}{2},\frac{m}{2},1,z).
\end{eqnarray}
For the reason of $a-b = a-(-a)=2a=-m\in\mathbb{Z}$, the regular solution cannot be written in a simple superposition of hyper-geometric function any more \cite{hyper}. Rather, it has the form of Eq.~(15.8.8) in Ref.~\cite{hyper}  
\begin{eqnarray}\label{15.8.8}
{}_2F_1\!\left(-\frac{m}{2},\frac{m}{2};1;z\right)
&=&
\frac{(-z)^{m/2}}{\Gamma\!\left(\tfrac{m}{2}\right)}
\sum_{k=0}^{m-1}
\frac{\left(-\tfrac{m}{2}\right)_k\,(m-k-1)!}
{k!\,\Gamma\!\left(1+\tfrac{m}{2}-k\right)}
\,z^{-k}
\\[6pt]\notag
&&+\frac{(-z)^{m/2}}{\Gamma\!\left(-\tfrac{m}{2}\right)}
\sum_{k=0}^{\infty}
\frac{\left(\tfrac{m}{2}\right)_k}
{k!\,(k+m)!\,\Gamma\!\left(1-\tfrac{m}{2}-k\right)}
(-1)^k\,z^{-k-m}
\\[6pt]\notag
&&\quad\times
\Big[\ln(-z)+\psi(k+1)+\psi(k+m+1)-\psi\!\left(\tfrac{m}{2}+k\right)
-\psi\!\left(1-\tfrac{m}{2}-k\right)\Big].
\end{eqnarray}
When only considering the dominant term $k=0$ and take $z\to\infty$, the solution has a simple form from which we can distinguish the growing branch and the decay branch
\begin{equation}
R_{1}(z)
\to C_{+}(m)\,(-z)^{m/2}
+(-z)^{-m/2}
[C_{-}(m)\,\ln(-z)+C_{0}(m)].
\end{equation}
It thus follows that the second term in eq.~\eqnref{15.8.8} plays the role of the decaying mode when evaluated at $k=0$.
We observe that when $m$ is even, this second term in eq.~\eqnref{15.8.8} develops a pole through the factor $\Gamma(-m/2)$. Hence the denominator diverges when $m$ is even, making this term vanishes identically. So the Bosonic Love number defined by 
\begin{eqnarray}
    \mathcal{F}_{m}=\frac{C_{-}(m)\,\ln(-z)+C_{0}(m)}{C_{+}(m)}
\end{eqnarray}
vanishes under this condition.
 In summary, for the $(2+1)$-dimensional acoustic black hole, the Bosonic Love number vanishes for even $m$, whereas for odd $m$ it acquires a logarithmic ratio structure. Although this structure is rather intricate, it is generically nonvanishing.

\subsection{Spinor field}
Finally, we explore the effective Dirac spinor field at the background Eq.~\eqnref{3dimmetric}. Note the Minkowski signature
\begin{equation}
\eta_{ab}=\mathrm{diag}(-,+,+).
\end{equation}
We choose the most natural orthonormal tetrad \cite{LiRen:2008BTZDiracTunnel}
\begin{equation}
e^{0}=c_s\sqrt{f}\,dt,\qquad
e^{1}=\frac{dr}{\sqrt{f}},\qquad
e^{2}=r\,d\phi .
\end{equation}
Then
\begin{equation}
ds^{2}=\eta_{ab}\,e^{a}e^{b}=-(e^{0})^{2}+(e^{1})^{2}+(e^{2})^{2}.
\end{equation}
The corresponding inverse frame (vector fields), satisfying
$e_{a}{}^{\mu}e^{b}{}_{\mu}=\delta_{a}{}^{b}$,
can be read off directly as
\begin{equation}
e_{0}{}^{\mu}=\left(\frac{1}{c_s\sqrt{f}},\,0,\,0\right),\qquad
e_{1}{}^{\mu}=\left(0,\,\sqrt{f},\,0\right),\qquad
e_{2}{}^{\mu}=\left(0,\,0,\,\frac{1}{r}\right).
\end{equation}
Use the torsion-free condition
\begin{equation}
de^{a}+\omega^{a}{}_{b}\wedge e^{b}=0,\qquad
\omega_{ab}=-\omega_{ba},
\end{equation}
where $de^{a}$ can be calculated
\begin{eqnarray}
de^{0}&=& c_s\, d(\sqrt{f}) \wedge dt= c_s\,\frac{f'}{2\sqrt{f}}\, dr \wedge dt, \\[6pt]
de^{1}&=& d\!\left(\frac{1}{\sqrt{f}}\right)\wedge dr= -\,\frac{f'}{2 f^{3/2}}\, dr \wedge dr = 0, \\[6pt]
de^{2}&=& dr \wedge d\phi .
\end{eqnarray}
Thus we obtain
\begin{eqnarray}
\omega^{01} = \frac{c_s}{2} f'(r)\, dt\,\to
\omega_{t}^{01} = \frac{c_s}{2}\, f',\\
\omega^{21}=\sqrt{f}\,d\phi=-
\omega^{12}
\to\omega_{\phi}^{12}=-\sqrt{f}.
\end{eqnarray}

Massless Dirac field $\Psi$ has two component, and satisfy equation in (2+1) dimension \cite{LiRen:2008BTZDiracTunnel}
\begin{eqnarray}
    i\hbar\,\gamma^{a} e_{a}{}^{\mu}\nabla_{\mu}\Psi= 0,
\end{eqnarray}
where
\begin{eqnarray}
\Psi=\begin{pmatrix}
\psi_{1} \\
\psi_{2}
\end{pmatrix},\qquad
\nabla_\mu = \partial_\mu + \frac{1}{4}\,\omega_\mu^{ab}\,\gamma_{[a}\gamma_{b]},\qquad
\gamma_{[a}\gamma_{b]} = \frac{1}{2}\bigl(\gamma_a\gamma_b - \gamma_b\gamma_a\bigr).
\end{eqnarray}
Note 
\begin{eqnarray}
    \gamma_0 = -\,\gamma^{0}=-i\sigma^{2}=\begin{pmatrix}
0 & -1 \\
1 & 0
\end{pmatrix}, \,\,
\gamma_1 = \gamma^{1}=\sigma^{1}=\begin{pmatrix}
0 & 1 \\
1 & 0
\end{pmatrix},\,\,
\gamma_2 = \gamma^{2}=\sigma^{3}=\begin{pmatrix}
1 & 0 \\
0 & -1
\end{pmatrix},\,\,
\end{eqnarray}
So we get the relation as following
\begin{eqnarray}
    \gamma^{a} e_{a}{}^{\mu}\,\partial_{\mu}
= \frac{\gamma^{0}}{c_s\sqrt{f}}\,\partial_t
+ \gamma^{1}\sqrt{f}\,\partial_r
+ \frac{\gamma^{2}}{r}\,\partial_\phi ,\\
\gamma^{a} e_{a}{}^{\mu}
\left( \frac{1}{4}\,\omega_{\mu}{}^{bc}\,\gamma_{[b}\gamma_{c]} \right)
=\frac{f'}{4\sqrt{f}}\;\gamma^{0}\gamma_{[0}\gamma_{1]} 
-\frac{\sqrt{f}}{2r}\;\gamma^{2}\gamma_{[1}\gamma_{2]} .
\end{eqnarray}
Using the relations, Dirac equation in (2+1) dimension can be written as
\begin{eqnarray}
\left(
\sqrt{f}\,\partial_r
+ \frac{f'}{8\sqrt{f}}
+ \frac{\sqrt{f}}{4r}
+ \frac{1}{c_s \sqrt{f}}\,\partial_t
\right)\psi_2
+ \frac{1}{r}\,\partial_\phi \psi_1
= 0 ,
\\[6pt]
\left(
\sqrt{f}\,\partial_r
+ \frac{f'}{8\sqrt{f}}
+ \frac{\sqrt{f}}{4r}
- \frac{1}{c_s \sqrt{f}}\,\partial_t
\right)\psi_1
- \frac{1}{r}\,\partial_\phi \psi_2
= 0 .
\end{eqnarray}
We now adopt the separation of variables ansatz
\begin{equation}
\psi_1(t,r,\phi)=R_1(r)\,e^{-i\omega t}\,e^{i m \phi},\qquad
\psi_2(t,r,\phi)=R_2(r)\,e^{-i\omega t}\,e^{i m \phi}.
\end{equation}
Then the radial equation in the limit $\omega\to0$ is 
\begin{eqnarray}
\sqrt{f}\,R_2'
+\left(\frac{f'}{4\sqrt{f}}+\frac{\sqrt{f}}{2r}
\right) R_2+i\,\frac{m}{r}\,R_1=0,
\\[6pt]
\sqrt{f}\,R_1'
+\left(\frac{f'}{4\sqrt{f}}+\frac{\sqrt{f}}{2r}
\right) R_1-i\,\frac{m}{r}\,R_2=0.
\end{eqnarray}
Note $m=0,\pm1,\pm2$...but we require $m$ positive for simplicity. It is the eigenvalue of the angular equation. To make the equation easier to explore, we take
\begin{equation}
R_{1,2}(r)=f(r)^{-1/4}\, r^{-1/2}\, u_{1,2}(r).
\end{equation}
So we rewrite the radial equation in the following simple form
\begin{eqnarray}
u_2'+ i\,\frac{m}{r\sqrt{f}}\,u_1= 0,\qquad
u_1'- i\,\frac{m}{r\sqrt{f}}\,u_2= 0.
\end{eqnarray}
Finally we get two identical radial equations of motion as
\begin{eqnarray}
u_1''+\left(\frac{1}{r}+\frac{f'}{2f}\right)u_1' -\frac{m^2}{r^2 f}\,u_1=0\qquad
u_2''+\left(\frac{1}{r}+\frac{f'}{2f}\right)u_2' -\frac{m^2}{r^2 f}\,u_2=0.
\end{eqnarray}
Next, we discard the subscripts of $u_{1,2}$ and denote both functions collectively by $u(r)$. Following the method we used in scalar field case, we substitute $f(r)=1-(r_{+}/r)^2$ and also choose a new variable
\begin{equation}
z = 1 - \left( \frac{r}{r_+} \right)^2 .
\end{equation}
Through it the radial equation can be expressed as a hyper-geometric form
\begin{equation}
z(1-z)\,\frac{d^2u}{dz^2}+\left(\frac{1}{2}-z\right)\frac{du}{dz}+\frac{m^2}{4}\,u=0,
\end{equation}
with $a=-m/2$,\,\,$b=m/2$ and $c=1/2$. The regular solution at horizon is
\begin{eqnarray}
    u(z)={}_2F_{1}(-\frac{m}{2},\frac{m}{2},\frac{1}{2},z).
\end{eqnarray}
Now because $a=-b$ and $c=1/2$, the function can be written in terms of elementary functions using Eq.~(15.4.11) in Ref.~\cite{hyper}
\begin{equation}
{}_2F_{1}\!\left(-\frac{m}{2},\frac{m}{2};\frac{1}{2};-x^{2}\right)
=\frac{1}{2}\left[\left(\sqrt{1+x^{2}}+x\right)^{m}
+\left(\sqrt{1+x^{2}}-x\right)^{m}\right].
\end{equation}
Replacing $x^2\to -z$ and take the limit $z\to -\infty$, we get
\begin{eqnarray}
    u(z)\to \frac{1}{2}[2^{m}(-z)^{\frac{m}{2}}+2^{-m}(-z)^{-\frac{m}{2}}].
\end{eqnarray}
Finally we obtain the Fermionic Love number 
\begin{eqnarray}
    \mathcal{F}_{m}=\frac{1}{4^m}.
\end{eqnarray}
It is therefore evident that, for Schwarzschild black holes as well as for three- and four-dimensional static acoustic black holes, the Fermionic Love numbers always take a power-law form and are generically nonvanishing.

\section{Conclusion}

In this work we investigated the static ($\omega\to 0$) tidal response of static acoustic black holes (ABHs) by explicitly computing both Bosonic ($s=0$) and Fermionic ($s=1/2$) Love numbers in (3+1) and (2+1) dimensions. The strategy was uniform across all cases: impose regularity at the acoustic horizon to select the physical branch of the solution, and expand the same solution at large radius to identify the decaying and growing modes whose coefficient ratio defines the dimensionless Love number.

For the (3+1)-dimensional ABH, the scalar sector generically yields a nonvanishing Love number (with vanishing only for special discrete $\ell$), reflecting the fact that an analogue black hole made of ordinary matter need not display the rigid-body-like behavior for scalar field familiar with Schwarzschild in general relativity. In contrast, the Fermionic Love number takes a remarkably simple and universal power-law form: for the two spin branches one finds
$F^{\pm 1/2}_{\ell m}=\pm 4^{-(\ell+1/2)}$. 

In $(2+1)$ dimensions, the scalar field exhibits a different mechanism: because $a-b\in\mathbb Z$ in the relevant hyper-geometric parameters, the large radius expansion contains logarithmic structures, so the decaying piece must be identified through the $\ln(-z)$ term. 
It is also shown that for even $m$ the decaying modes is eliminated by the $\Gamma(-m/2)$ pole, so the Bosonic Love number vanishes, while for odd $m$ Love numer survives as a nontrivial logarithmic-ratio structure. By comparison, the (2+1)-dimensional Fermionic Love number again remains a simple power law, $F_m=4^{-m}$, and is generically nonzero.

These results provide a concrete benchmark for how black-hole-like response properties do (and do not) persist in analogue systems, and they sharpen the contrast between integer-spin and half-integer-spin sectors. Natural extensions include incorporating rotation
, going beyond $\omega=0$ to the full frequency-dependent retarded response, and exploring the underlying reason of non-zero Fermionic Love number beyond symmetry considerations.
\begin{acknowledgements}
		This work is supported by National Natural Science Foundation of China (NSFC) with Grant No. 12275087. 
	\end{acknowledgements}

    \newpage
	\par
	\bibliographystyle{unsrt}

\end{document}